\documentclass[12pt]{article}
\begin{document}
\pagestyle{plain}
\setcounter{page}{1}
\begin{center}
{\large\bf Non-Anticommutative Quantum Gravity}
\vskip 0.3 true in
{\large J. W. Moffat}
\vskip 0.3 true in
{\it Department of Physics, University of Toronto,
Toronto, Ontario M5S 1A7, Canada}
\end{center}
\begin{abstract}%
A calculation of
the one loop gravitational self-energy graph in non-anticommutative quantum
gravity reveals that graviton loops are damped by internal momentum dependent factors
in the modified propagator and the vertex functions. The non-anticommutative quantum
gravity perturbation theory is finite for matter-free gravity and
for matter interactions.
\end{abstract}



\section{Introduction}

Recently, the consequences for
perturbative quantum gravity were investigated, when the gravitational action is
given on a noncommutative spacetime geometry, by expanding the
metric about a flat Minkowski spacetime and by taking the usual Einstein-Hilbert
action, whose fields are functions on ordinary commutative spacetime, except that
the products of field quantities are formed by using the Moyal $\star$-product
rule~\cite{Moffat}. The first order, one loop graviton self-energy was calculated,
using a noncommutative action and functional generator $Z[j_{\mu\nu}]$. It was shown
that the planar one loop graviton graph and vacuum polarization are essentially the
same as for the commutative perturbative result, while the non-planar graviton loop
graphs were damped due to the oscillatory behavior of the noncommutative phase factor
in the Feynman integrand. Thus, the overall noncommutative perturbative theory
remained unrenormalizable and divergent.

In the following, we shall repeat the analysis of perturbative quantum gravity by
using a non-anticommutative geometry, defined by the superspace
coordinates~\cite{Moffat2}:
\begin{equation}
\rho^\mu=x^\mu+\beta^\mu,
\end{equation}
where the $x^\mu$ denote the classical commuting c-number coordinates of spacetime,
and $\beta^\mu$ denote Grassman coordinates which satisfy
\begin{equation}
\{\beta^\mu,\beta^\nu\}=\beta^\mu\beta^\nu+\beta^\nu\beta^\mu=0.
\end{equation}
The familiar commutative coordinates of spacetime
are replaced by operators ${\hat\rho}^\mu$, which satisfy
\begin{equation}
\{{\hat\rho}^\mu,{\hat\rho}^\nu\}=2x^\mu x^\nu+2(x^\mu\beta^\nu
+x^\nu\beta^\mu)-\tau^{\mu\nu},
\end{equation}
where $\tau^{\mu\nu}$ is a symmetric two-tensor.

In the non-anticommutative field theory formalism,
the product of two operators ${\hat f}$ and ${\hat g}$ has a corresponding
$\diamondsuit$-product
\begin{equation}
({\hat f}\diamondsuit {\hat g})(\rho)
=\exp\biggr(-\frac{1}{2}\tau^{\mu\nu}\frac{\partial}{\partial\xi^\mu}\frac{\partial}
{\partial\zeta^\nu}\biggr)f(\rho+\xi)g(\rho+\zeta)\vert_{\xi=\zeta=0}
$$ $$
=f(\rho)g(\rho)-\frac{1}{2}\tau^{\mu\nu}\partial_\mu f(\rho)\partial_\nu
g(\rho)+O(\tau^2),
\end{equation}
where $\partial_\mu=\partial/\partial\rho_\mu$.

In Section 2, we shall expand the metric tensor of general relativity about flat
Minkowskian spacetime and replace all commutative products of gravitational fields
and their derivatives in the Einstein-Hilbert action by $\diamondsuit$-products
in the superspace with coordinates $\rho^\mu$. In Section 3, a calculation of the
first order, self-energy graviton loop diagrams reveals that they are finite due to
the damping of the ultraviolet divergences by the modified graviton propagator. This
finite behavior of the loop amplitudes holds for all higher order diagrams. A
discussion of the nonlocal nature of the non-anticommutative quantum gravity
formalism and the unitarity of graviton amplitudes is given in Section 4, and
concluding remarks are made in Section 5.

\section{\bf The Non-Anticommutative Gravity Action}

We define the non-anticommutative gravitational action in the superspace manifold as
\begin{equation}
S_{\rm grav}=-\frac{2}{\kappa^2}\int d^4\rho(\sqrt{-g}\diamondsuit
R+2\sqrt{-g}\lambda),
\end{equation}
where we use the notation: $\mu,\nu=0,1,2,3$,
$g={\rm det}(g_{\mu\nu})$, the metric signature of Minkowski spacetime is
$\eta_{\mu\nu}={\rm diag}(1,-1,-1,-1)$, $R=g^{\mu\nu}\diamondsuit R_{\mu\nu}$ denotes
the scalar curvature, $\lambda$ is the cosmological constant and $\kappa^2=32\pi G$
with $c=1$. The Riemann tensor is defined such that
\begin{equation}
{R^\lambda}_{\mu\nu\rho}=\partial_\rho{\Gamma_{\mu\nu}}^\lambda
-\partial_\nu{\Gamma_{\mu\rho}}^\lambda+
{\Gamma_{\mu\nu}}^\alpha\diamondsuit{\Gamma_{\rho\alpha}}^\lambda
-{\Gamma_{\mu\rho}}^\alpha\diamondsuit{\Gamma_{\nu\alpha}}^\lambda.
\end{equation}

The gravitational action $S_{\rm grav}$ in our non-anticommutative geometry
can be rewritten as
\begin{equation}
\label{action}
S_{\rm grav}=\frac{1}{2\kappa^2}\int
d^4\rho [({\bf g}^{\rho\sigma}\diamondsuit{\bf g}_{\lambda\mu}\diamondsuit{\bf
g}_{\kappa\nu} $$ $$ -\frac{1}{2}{\bf g}^{\rho\sigma} \diamondsuit{\bf
g}_{\mu\kappa}\diamondsuit{\bf g}_{\lambda\nu}
-2\delta^\sigma_\kappa\delta^\rho_\lambda{\bf
g}_{\mu\nu})\diamondsuit\partial_\rho{\bf
g}^{\mu\kappa}\diamondsuit\partial_\sigma{\bf g}^{\lambda\nu}],
\end{equation}
where ${\bf g}^{\mu\nu}=\sqrt{-g}g^{\mu\nu}$ and in the following we shall omit
the cosmological constant $\lambda$.
We expand the local interpolating graviton field ${\bf g}^{\mu\nu}$ as
\begin{equation}
{\bf g}^{\mu\nu}=\eta^{\mu\nu}+\kappa\gamma^{\mu\nu}+O(\kappa^2).
\end{equation}
Then, for the non-anticommutative superspace
\begin{equation}
{\bf g}_{\mu\nu}=\eta_{\mu\nu}-\kappa\gamma_{\mu\nu}
+\kappa^2{\gamma_\mu}^\alpha\diamondsuit{\gamma_\alpha}_\nu
-\kappa^3{\gamma_\mu}^\alpha
\diamondsuit\gamma_{\alpha_\beta}\diamondsuit{\gamma_\nu}^\beta+O(\kappa^4).
\end{equation}

Let us consider the non-anticommutative generating
functional~\cite{Fradkin,Leibbrandt,Medrano}, defined in the superspace manifold:
\begin{equation}
Z[j_{\mu\nu}]=\int
d[{\bf g}^{\mu\nu}]\Delta[{\bf g}^{\mu\nu}] \exp i\biggl[S_{\rm
grav}+\frac{1}{\kappa}\int d^4\rho{\bf g}^{\mu\nu}\diamondsuit j_{\mu\nu}
$$ $$
-\frac{1}{\kappa^2\beta}\int d^4\rho\partial_\mu{\bf g}^{\mu\nu}
\diamondsuit\partial_\alpha{\bf g}^{\alpha\beta}\eta_{\nu\beta}\biggr],
\end{equation}
where
$(\partial_\mu{\bf g}^{\mu\nu}\diamondsuit\partial_\alpha{\bf
g}^{\alpha\beta}\eta_{\nu\beta}) /\kappa^2\beta$ is the gauge fixing term.
Moreover, $\Delta$ can be interpreted in terms of fictitious particles and is
given by
\begin{equation}
\Delta[{\bf g}^{\mu\nu}]^{-1} =\int
d[\xi_\lambda]d[\eta_\nu]\exp i\biggl\{\int
d^4\rho\eta^\nu\diamondsuit[\eta_{\nu\lambda}
\partial^\sigma\partial_\sigma-\kappa(\partial_\lambda\partial^\mu
\gamma_{\mu\nu}-\gamma_{\mu\rho}
$$ $$
\times\eta_{\nu\lambda}\partial^\mu\partial^\rho-\partial^\mu\gamma_{\mu\rho}
\eta_{\nu\lambda}
\partial^\rho+\partial^\mu\gamma_{\mu\nu}\partial_\lambda)]\diamondsuit
\xi^\lambda\biggr\},
\end{equation}
where $\xi^\lambda$ and $\eta^\lambda$ are the fictitious ghost
particle fields.

The gravitational action is expanded as
\begin{equation}
S_{\rm grav}=S^{(0)}_{\rm grav}+\kappa S^{(1)}_{\rm grav}+\kappa^2
S^{(2)}_{\rm grav}+....
\end{equation}
We find the following expanded values of $S_{\rm grav}$:
\begin{equation}
S^{(0)}_{\rm grav}=\int d^4\rho(\frac{1}{2}\partial_\sigma\gamma_{\lambda\rho}
\diamondsuit\partial^\sigma\gamma^{\lambda\rho}-\partial_\lambda\gamma^{\rho\kappa}
\diamondsuit\partial_\kappa\gamma^\lambda_\rho-\frac{1}{4}\partial_\rho\gamma
\diamondsuit\partial^\rho\gamma
$$ $$
-\frac{1}{\alpha}\partial_\rho\gamma^\rho_\lambda\diamondsuit\partial_\kappa
\gamma^{\kappa\lambda}),
\end{equation}
\begin{equation}
S^{(1)}_{\rm grav}=\frac{1}{4}\int
d^4\rho(-4\gamma_{\lambda\mu}\diamondsuit\partial^\rho\gamma^{\mu\kappa}
\diamondsuit\partial_\rho{\gamma_\kappa}^\lambda+2\gamma_{\mu\kappa}
\diamondsuit\partial^\rho\gamma^{\mu\kappa}\diamondsuit\partial_\rho\gamma $$ $$
+2\gamma^{\rho\sigma}\diamondsuit\partial_\rho\gamma_{\lambda\nu}
\diamondsuit\partial_\sigma\gamma^{\lambda\nu}
-\gamma^{\rho\sigma}\diamondsuit\partial_\rho\gamma\diamondsuit\partial_\sigma\gamma
+4\gamma_{\mu\nu}\diamondsuit\partial_\lambda\gamma^{\mu\kappa}
\diamondsuit\partial_\kappa\gamma^{\nu\lambda}),
\end{equation}
\begin{equation}
S^{(2)}_{\rm grav}=\frac{1}{4}\int d^4\rho[4\gamma_{\kappa\alpha}
\diamondsuit\gamma^{\alpha\nu}
\diamondsuit\partial^\rho\gamma^{\lambda\kappa}\diamondsuit\partial_\rho\gamma_{\nu\lambda}
+(2\gamma_{\lambda\mu}\diamondsuit\gamma_{\kappa\nu}
-\gamma_{\mu\kappa}\diamondsuit\gamma_{\nu\lambda})
\diamondsuit\partial^\rho\gamma^{\mu\kappa}\diamondsuit\partial_\rho\gamma^{\nu\lambda}
$$ $$
-2\gamma_{\lambda\alpha}\diamondsuit{\gamma_\nu}^\alpha\diamondsuit\partial^\rho\gamma^{\lambda\nu}
\diamondsuit\partial_\rho\gamma-2\gamma^{\rho\sigma}\diamondsuit{\gamma_\nu}^\kappa
\diamondsuit\partial_\rho\gamma_{\lambda\kappa}\diamondsuit\partial_\sigma\gamma^{\nu\lambda}
$$ $$
+\gamma^{\rho\sigma}\diamondsuit\gamma^{\nu\lambda}\diamondsuit\partial_\sigma\gamma_{\nu\lambda}
\diamondsuit\partial_\rho\gamma-2\gamma_{\mu\alpha}\diamondsuit\gamma^{\alpha\nu}
\diamondsuit\partial^\lambda\gamma^{\mu\kappa}\diamondsuit\partial_\kappa\gamma_{\nu\lambda}],
\end{equation}
where $\gamma={\gamma^\alpha}_\alpha$.

The free part of the action is not the same as the commutative case. However, we
can choose to quantize the field quantities $\gamma_{\mu\nu}$ with the same vacuum
state as in the commutative case~\cite{Moffat2}. In particular, the measure
in the functional integral formalism is the same as the commutative theory, for in
momentum space the additional factors due to the use of the $\diamondsuit$-product
disappear when we impose the normalization condition for the partition function.

The modified graviton propagator in the fixed
gauge $\beta=-1$ is given in superspace by~\cite{Moffat2}
\begin{equation}
i{\bar D}^{\rm grav}_{\mu\nu\rho\sigma}(\rho-\rho')
=(\eta_{\mu\rho}\eta_{\nu\sigma}+\eta_{\mu\sigma}\eta_{\nu\rho}
-\eta_{\mu\nu}\eta_{\rho\sigma})
$$ $$
\times\frac{i}{(2\pi)^4}\int
\frac{d^4p\exp[\frac{1}{2}(p\tau p)]}{p^2+i\epsilon}\exp[ip(\rho-\rho')],
\end{equation}
where $(p\tau p)=p_\mu\tau^{\mu\nu}p_\nu$. In momentum space this becomes
\begin{equation}
i{\bar D}^{\rm grav}_{\mu\nu\rho\sigma}(p)
=(\eta_{\mu\rho}\eta_{\nu\sigma}+\eta_{\mu\sigma}\eta_{\nu\rho}
-\eta_{\mu\nu}\eta_{\rho\sigma})\frac{i\exp[\frac{1}{2}(p\tau p)]}{p^2+i\epsilon}.
\end{equation}
The ghost propagator in momentum space is
\begin{equation}
i{\bar D}^G_{\mu\nu}(p)=\frac{\eta_{\mu\nu}i\exp[\frac{1}{2}(p\tau
p)]}{p^2+i\epsilon}.
\end{equation}

In momentum space, a graviton interaction diagram has
an additional factor which takes the form~\cite{Moffat2}:
\begin{equation}
V(q_1,q_2...,q_n)=\exp\biggl(\frac{1}{2}\sum_{i<j}q_i\cdot q_j\biggr),
\end{equation}
where
\begin{equation}
q_i\cdot q_j\equiv q_{i\mu}\tau^{\mu\nu}q_{j\nu}.
\end{equation}
In flat spacetime, the only changes of the Feynman rules consist of
inserting a modified graviton propagator ${\bar D}^{\rm grav} _{\mu\nu\alpha\beta}$
in every internal line and inserting a factor $V(q_1,...,q_n)$ in every diagram.

Let us consider the effects of an infinitesimal gauge transformation
\begin{equation}
\rho^{'\mu}=\rho^\mu+\zeta^\mu
\end{equation}
on the non-anticommutative generating functional $Z[j_{\mu\nu}]$, where $\zeta^\mu$
can depend on $\rho^\mu$ and $\gamma^{\mu\nu}$. We get
\begin{equation}
\delta{\bf g}^{\mu\nu}(\rho)
=-\zeta^\lambda(\rho)\diamondsuit\partial_\lambda{\bf g}^{\mu\nu}(\rho)
+\partial_\rho\zeta^\mu(\rho)\diamondsuit{\bf g}^{\rho\nu}(\rho) $$ $$
+\partial_\sigma\zeta^\nu(\rho)\diamondsuit{\bf g}^{\mu\sigma}(\rho)
-\partial_\alpha\zeta^\alpha(\rho)\diamondsuit{\bf g}^{\mu\nu}(\rho).
\end{equation}
We now find that
\begin{equation}
\label{gaugetransf}
\delta\gamma_{\mu\nu}(\rho)=-\zeta^\lambda\diamondsuit\partial_\lambda\gamma_{\mu\nu}
+\partial^\rho\zeta_\mu\diamondsuit\gamma_{\rho\nu}+\partial^\rho\zeta_\nu\diamondsuit\gamma_{\mu\rho}
$$ $$ -\partial^\rho\zeta_\rho\diamondsuit\gamma_{\mu\nu}
+\frac{1}{\kappa}(\partial_\nu\zeta_\mu+\partial_\mu\zeta_\nu-\partial^\rho\zeta_\rho\eta_{\mu\nu}).
\end{equation}
The functional generator $Z$ is invariant under changes in the
integration variable and the transformation $(\ref{gaugetransf})$.

The non-anticommutative gauge transformations (\ref{gaugetransf}) should be
considered part of an $NACSO(3,1)$ group of gauge transformations. It is clear
that in the limit $\beta^\mu\rightarrow 0$ and $\vert\tau^{\mu\nu}\vert\rightarrow
0$, the standard local Lorentz group of gauge transformations $SO(3,1)$ is recovered.

\section{Gravitational Self-Energy}

The lowest order contributions to the graviton self-energy will include
the standard graviton loops, the ghost field loop contributions and the measure loop
contributions. In perturbative gravity theory, the first order vacuum
polarization tensor $\Pi^{\mu\nu\rho\sigma}$ must satisfy the Slavnov-Ward
identities~\cite{Medrano}:
\begin{equation}
\label{SlavnovWard}
p_\mu p_\rho
{\bar D}^{\mu\nu\alpha\beta}(p)\Pi_{\alpha\beta\gamma\delta}(p)
{\bar D}^{\gamma\delta\rho\sigma}(p)=0.
\end{equation}

The basic lowest order graviton self-energy diagram is determined by
\begin{equation}
\Pi_{\mu\nu\rho\sigma}(p)
=\frac{1}{2}\kappa^2
\int d^4q {\cal U}_{\mu\nu\alpha\beta\gamma\delta}(p,-q,q-p)
{\bar D}^{{\rm grav}\,\alpha\beta\kappa\lambda}(q)
$$ $$
\times{\bar D}^{{\rm grav}\,\gamma\delta\tau\xi}(p-q)
{\cal U}_{\kappa\lambda\tau\xi\rho\sigma}(q,p-q,-p)V(q,p-q,p),
\end{equation}
where ${\cal U}$ is the three-graviton vertex function
\begin{equation}
{\cal U}_{\mu\nu\rho\sigma\delta\tau}(q_1,q_2,q_3) =
-\frac{1}{2}\biggl[q_{2(\mu}q_{3\nu)}\biggl(2\eta_{\rho(\delta}\eta_{\tau)\sigma}
-\eta_{\rho\sigma}\eta_{\delta\tau}\biggr)
$$ $$
+q_{1(\rho}q_{3\sigma)}\biggl(2\eta_{\mu(\delta}\eta_{\tau)\nu}
-\eta_{\mu\nu}\eta_{\delta\tau}\biggr)+...\biggr],
\end{equation}
and the ellipsis denote similar contributions. We must add to this result
the contributions from the one loop fictitious ghost particle graph
and tadpole graph.

The modified graviton propagator has the asymptotic behavior
as $q^2\rightarrow\infty$:
\begin{equation}
{\bar D}^{\rm grav}\sim \exp\biggl[\frac{1}{2}(q\tau q)\biggr].
\end{equation}
If we choose an orthonormal frame such that
$\tau^{\mu\nu}=\eta^{\mu\nu}/\Lambda_{\rm grav}^2$, then we get~\cite{Moffat2}
\begin{equation}
{\bar D}^{\rm grav}\sim \exp\biggl(\frac{1}{2}q^2/\Lambda^2_{\rm grav}\biggr).
\end{equation}
We can now perform an analytic continuation in the invariant momentum $q$ such
that $q^2=-k^2$ with $k^2=k_4^2+k_1^2+k_2^2+k_3^2 >0$. Then, we have in Euclidean
momentum space
\begin{equation}
{\bar D}^{\rm grav}\sim
\exp\biggl(-\frac{1}{2}k^2/\Lambda^2_{\rm grav}\biggr).
\end{equation}

The asymptotic behavior in Euclidean momentum space of the modified graviton
propagator will damp out the ultraviolet behavior of the integrands in the graviton
one loop self-energy diagrams. Thus, these diagrams lead to a finite lowest
order vacuum polarization in non-anticommutative quantum gravity. Higher order
graviton self-energy loops will contain products of the modified propagator,
corresponding to the number of internal lines in a loop diagram, which will also damp
out the ultraviolet behavior of the internal momentum integrations. It follows that
the higher order pure gravity loops and loops occurring in gravity-matter
interactions will be finite.

In order to retain physical behavior of graviton scattering amplitudes and crossing
symmetry relations, and avoid essential singularities at infinite momentum, we are
required to choose an orthonormal frame with $\tau^{00}=\tau^{0n}=0$ and
$\tau^{mn}=-\delta^{mn}/\Lambda^2_{\rm grav}\quad (m,n=1,2,3)$~\cite{Moffat3}. It
follows that for ${\bf q}^2 >0$ (where ${\bf q}$ denotes the three-momentum vector)
, we have for $q^2\rightarrow \pm\infty$:
\begin{equation}
{\bar D}^{\rm grav}\sim
\exp\biggl(-\frac{1}{2}{\bf q}^2/\Lambda^2_{\rm grav}\biggr).
\end{equation}

\section{Nonlocality and Unitarity}

The infinite derivatives that occur in the $\diamondsuit$-product of fields render
the non-anticommutative field theories nonlocal. This is also true for
noncommutative quantum field theory~\cite{Minwalla,Seiberg,Seiberg2} and
noncommutative quantum gravity~\cite{Moffat,Moffat4}. However, with the choice
$\tau^{00}=\tau^{n0}=0, \tau^{mn}\not= 0$, we can suppress some of the potentially
unphysical acausal behavior of the non-anticommutative amplitudes and retain some of
the standard features of quantum field theory such as the canonical Hamiltonian
formalism. But if we consider a non-perturbative treatment of quantum gravity in a
non-anticommutative geometry, then standard perturbative field theory methods do not
apply and it is possible that one must view the nonlocal dynamics in a new way. Only
future developments in non-perturbative quantum gravity, and, in particular, the way
such developments will affect calculations using non-anticommutative geometry, can
provide us with deeper insights into these problems.

An analysis of the unitarity of amplitudes in scalar non-anticommutative field
theory~\cite{Moffat3}, showed that it can be satisfied, because the amplitudes in
non-anticommutative field theory are only modified by {\it entire} functions, which
do not introduce any new unphysical singularities in a finite region of the complex
momentum plane. Moreover, with the choice $\tau^{00}=\tau^{n0}=0, \tau^{mn}\not= 0$,
the scattering amplitudes are regular at infinite energies and crossing symmetry
relations retain their physical behavior. These results can be extended to
non-anticommutative quantum gravity. This choice of the symmetric tensor
$\tau^{\mu\nu}$ breaks local Lorentz invariance. However, it is possible to break the
Lorentz symmetry `softly' by a spontaneous symmetry breaking mechanism, which
involves adding a Higgs breaking mechanism contribution to the action that uses a
Higgs vector to break local $SO(3,1)$ to $O(3)$ at the small distance $\ell_{\rm
grav}$ ($\ell_{\rm grav}\sim 1/\Lambda_{\rm grav}$) when quantum gravity is
expected to become important~\cite{Moffat5}.

\section{\bf Conclusions}

We have developed a perturbative, non-anticommutative quantum gravity formalism
by using the $\diamondsuit$-product in the gravity action, wherever products
of gravitational fields and their derivatives occur in the superspace manifold. By
expanding about Minkowski flat spacetime, we were able to calculate the loop graphs
to first order. We find that by using the Feynman rules appropriate for
non-anticommutative quantum field theory, the loop graphs are finite to first order
and to all orders due to the Gaussian damping of the modified graviton propagator.
Since the loop graphs are finite, we do not find any peculiar ultraviolet/infrared
behavior as is encountered in noncommutative field theory~\cite{Minwalla}.

From these results, we expect that non-anticommutative Yang-Mills gauge theories
with the action defined in superspace
\begin{equation}
S_{\rm YM}=-\frac{1}{4}\int d^4\rho F^{a\mu\nu}\diamondsuit F_{a\mu\nu},
\end {equation}
where
\begin{equation}
F^a_{\mu\nu}=\partial_\nu A^a_\mu-\partial_\mu A^a_\nu
-\epsilon^{abc}A^b_\mu\diamondsuit A^c_\nu,
\end{equation}
will also yield finite loop diagrams with an energy scale parameter
$\Lambda_{\rm YM}$.

\vskip 0.2
true in {\bf Acknowledgments}
\vskip 0.2 true in
This work was supported by the Natural Sciences and Engineering Research Council of
Canada.
\vskip 0.5 true in

\end{document}